# Biaxial strain tuning of exciton energy and polarization in monolayer $WS_2$

G. Kourmoulakis[1,2†], A. Michail[3,4†], I. Paradisanos[1], X. Marie[5], M.M. Glazov[6], B. Jorissen[7], L. Covaci[7], E. Stratakis[1,8], K. Papagelis[4,9], J. Parthenios[4*], and G. Kioseoglou[1,2*]

[1]*Institute of Electronic Structure and Laser, Foundation for Research and Technology-Hellas, Heraklion, 70013, Greece*

[2]*Department of Materials Science and Technology, University of Crete, Heraklion, 70013, Greece*

[3]*Department of Physics, University of Patras, Patras, 26504, Greece*

[4]*Institute of Chemical Engineering Sciences, Foundation for Research and Technology-Hellas, Stadiou str Platani, Patras, 26504, Greece*

[5]*Universite de Toulouse, INSA-CNRS-UPS, LPCNO, 135 Av. Rangueil, 31077, Toulouse, France*

[6]*Ioffe Institute, St.-Petersburg, 194021, Russia*

[7]*University of Antwerp, Groenenborgerlaan 171, B-2020 Antwerpen, Belgium*

[8]*Department of Physics, University of Crete, Heraklion Crete 71003, Greece*

[9]*School of Physics, Dept. of Solid-State Physics, Aristotle University of Thessaloniki, Thessaloniki, 54124, Greece*

## ABSTRACT

We perform micro-photoluminescence and Raman experiments to examine the impact of biaxial tensile strain on the optical properties of $WS_2$ monolayers. A strong shift on the order of -130 meV per % of strain is observed in the neutral exciton emission at room temperature. Under near-resonant excitation we measure a monotonic decrease in the circular polarization degree under applied strain. We experimentally separate the effect of the strain-induced energy detuning and evaluate the pure effect coming from biaxial strain. The analysis shows that the suppression of the circular polarization degree under biaxial strain is related to an interplay of energy and polarization relaxation channels as well as to variations in the exciton oscillator strength affecting the long-range exchange interaction.

† These authors contributed equally

* Corresponding authors: jparthen@iceht.forth.gr, gnk@materials.uoc.gr

Monolayer (1L) transition metal dichalcogenides (TMDs) are direct gap semiconductors with strong spin-orbit interaction and unique optical[1,2] and mechanical properties, highly attractive for flexible photonic[3,4] and optoelectronic applications[5,6]. Their honeycomb lattice structure, combined with orbital hybridization, broken inversion and time-reversal symmetry make them potential candidates for valleytronics, a concept where the valley index is a potential new degree of freedom to store, manipulate and read out information[5,7,8]. Importantly, the optical transitions in 1L-TMDs are chiral as right and left-handed circularly polarized light can induce transitions only at the K and K′ valleys[9–12]. Thus, an imbalance in the carrier population between the two valleys can be optically or electrically generated, referred to as valley polarization (VP)[13]. Development of a fundamental understanding of various external perturbations affecting VP is a topical problem nowadays. For instance, excitation energy and temperature[14], dielectric environment[15,16], as well as strain[17] can influence the degree of VP. The role of mechanical strain in the optical properties of TMDs is a key point for operation of flexible optoelectronic devices[3,4]. It has important consequences as it can induce contrasting modulations in the symmetry and electronic states of the material, depending on the type of strain whether it is tensile or compressive, isotropic or anisotropic. Recently, the strain tuning of energy levels has been reported in 1L-$WS_2$ grown by chemical vapor deposition (CVD) upon application of isotropic, biaxial tensile strain[18]. A transition from indirect-to-direct band gap in $MoTe_2$ bilayers subjected to uniaxial strain has also been claimed lately[19]. While the dependence of the VP on the uniaxial strain has been studied in $MoS_2$ monolayers and bilayers[17,20], the impact of biaxial strain on the degree of VP in TMD monolayers is scarcely studied.

Here we apply up to 0.45% isotropic biaxial tensile strain in $WS_2$ monolayers and monitor the neutral exciton emission at room temperature. We observe a red shift of -130meV per percent of applied biaxial strain. To examine the impact of biaxial strain on the optical orientation of 1L-$WS_2$, we perform helicity-resolved photoluminescence experiments as a function of the applied strain and observe a drastic decrease in the degree of VP. We uncover that the VP is reduced due to both the strain-induced increase in the detuning between the excitation energy and the exciton resonance and the strain effect on the band structure and excitonic states. We identify two main effects underlying the reduction of the VP: (i) a strain-modified exciton oscillator strength, which in turn affects the long-range exchange interaction of electrons and holes and (ii) a suppressed K-Λ intervalley scattering channel leading to larger depolarization rates.

Bulk crystals of $WS_2$ and graphite (2D Semiconductors) are exfoliated using scotch tape (Nitto) and directly transferred on polydimethylsiloxane (PDMS) films (TELTEC) for inspection under the optical microscope. 1L-$WS_2$ and flat bulk flakes of graphite are identified and subsequently transferred on Si/$SiO_2$ (285nm) substrates for fabrication of graphite/1L-$WS_2$ heterostructures using a deterministic dry transfer protocol[21]. The strategic role of the graphite flake is to provide a sufficiently clean and flat support to 1L-$WS_2$ and to efficiently filter exciton complexes beyond neutral excitons via rapid charge and energy transfer processes (more details can be found in refs[15,22]). In addition, this process assists in sustaining large degrees of VP at room temperature[23]. Thermal annealing at 150ºC is applied for 20 minutes to improve the quality of the interface between the flakes. Finally, the complete heterostructure is encapsulated in poly-methyl methacrylate (PMMA, Microchem) and transferred on top of an elastic substrate with a cruciform shape (see detailed methodology in Supplementary Material, section A, Figure S1). The latter, is mounted on a custom device, designed to induce biaxial tensile strain (Figure 1a, b). For isotropic biaxial strain, $u_{xx} = u_{yy}$ must hold, where $u$ is the strain tensor and the subscript corresponds to the Cartesian coordinates (Figure 1b). To satisfy the condition $u_{xx} = u_{yy}$, the heterostructure is carefully placed close to the center of the cruciform (see details in Supplementary Material, section A, Figure S2). Room temperature photoluminescence (PL), differential reflectivity ($\Delta R = \frac{R_{on}-R_{off}}{R_{off}}$, with $R_{on}$ the intensity reflection coefficient of the sample and $R_{off}$ is the same structure without $WS_2$) and Raman spectroscopy are employed to evaluate the quality of the heterostructure and optically explore the impact of strain.

In Figure 2a we compare PL and $\Delta R$ spectra collected at T = 300 K between 1L-$WS_2$ on $SiO_2$/Si and 1L-$WS_2$ on graphite. A red shift on the order of 30 meV is observed in PL and $\Delta R$ of 1L-$WS_2$/graphite due to a local dielectric screening resulting in a reduction of the bandgap and exciton binding energy[24]. In addition, the emission intensity is suppressed roughly by one order of magnitude (Supplementary Material, section B, Figure S3) due to a rapid charge and energy transfer from 1L-$WS_2$ to graphite[22]. The good agreement between the PL emission energy and reflectivity spectra close to 2 eV suggests that the single emission peak observed in 1L-$WS_2$ on graphite originates from neutral excitons, in line with previous reports[15,22]. We further observe a symmetric PL emission with a 20 meV narrower linewidth in 1L-$WS_2$ on graphite, possibly due to filtering of the trion emission. Raman spectra of the heterostructure are presented in Figure 2b. The 60cm$^{-1}$ frequency difference between E′ and $A_1$′ vibrational modes further verifies the $WS_2$ monolayer thickness. In addition, the G and the deconvoluted in two components 2D mode confirm the presence of the underlying multi-layered graphite flake[25]. Near-resonant (2.087 eV) helicity-resolved PL experiments demonstrate a high VP degree of 1L-$WS_2$ on graphite (see also Supplementary Material, section B, Figure S4). We measure an average VP of ~25% by exciting with $\sigma^+$ light and detecting the PL emission intensity of both $\sigma^+$ and $\sigma^-$ components, with VP = $\frac{I^{\sigma^+}-I^{\sigma^-}}{I^{\sigma^+}+I^{\sigma^-}}$.

We now describe the experimental results in the presence of the biaxial strain. Verified by PL and Raman spectroscopy, the encapsulation of 1L-$WS_2$/graphite in PMMA layers mechanically supports the structure and efficiently transfers the applied stress on the cruciform to 1L-$WS_2$, yielding reproducible and reversible results (Supplementary Material, section C, Figure S5). In Figure 3, we plot helicity-resolved PL emission spectra as a function of tensile, isotropic biaxial strain. A clear redshift is observed with increasing biaxial tensile strain due to a reduction of the band gap (Figures 3 and 4a)[18]. We increase the strain up to 0.45% and we measure a total redshift of ~50meV in the emission energy, suggesting a rate of -130meV per % of applied strain in this range (Figure 4a). This value demonstrates very good agreement with ab initio calculations[26] showing a rate of ~ -133meV per % (see also Supplementary Material, section E, Figure S6d, for DFT calculations without considering excitonic effects). Interestingly, as strain increases, a monotonic decrease in the VP degree is observed while the value of VP is fully reversible under several strain cycles (Figure 4b).

Tunability of the degree of VP in monolayer TMDs has been reported before for uniaxial tensile strain[17] and more recently for small compressive biaxial strain as well[27]. Our target here is to understand the different microscopic contributions in the observed drop of the VP degree with increasing tensile biaxial strain (Figure 4b). First, we aim to distinguish the effect of the excess energy on the measured VP degree, which has not been thoroughly considered in the literature yet. As the excitation energy is fixed at 2.087 eV, the energy detuning increases between excitation and emission for every step of applied strain because of the red-shifted exciton emission. This rises the exciton's effective thermalization time, thus an exciton will eventually lose polarization as its energy will relax slower compared to the spin relaxation time before radiative recombination[28]. An additional effect that can contribute to the drop of the circular polarization degree of excitons by increasing the excess energy is related to the band nonparabolicity in TMD monolayers and violation of the chiral selection rules away from K and K' points (see detailed discussion in ref.[29]). However, this effect is too small to account for the detuning energy range studied here, see Supplementary Material section D for further theoretical analysis.

To evaluate the contribution of the excess energy variation we measure the VP by performing excitation energy dependent experiments in unstrained 1L-$WS_2$/graphite samples and we compare the results with the strain-dependent experiments. It is evident that the degree of VP decreases with a slope of (-0.17±0.02) %/meV (red spheres in Figure 5) which is smaller, in absolute value, than the slope of (-0.25±0.01) %/meV where the excess energy is introduced by biaxial strain (blue spheres in Figure 5, extracted from Figure 3). For the red spheres of Figure 5, the x-axis corresponds to the energy difference between the excitation laser energy and the value of 2.087 eV. For the blue spheres it corresponds to the excess energy introduced by the strain-shifted exciton emission under the fixed excitation energy (2.087 eV).

A comparison between the two slopes highlights that ≈70% of the observed VP degree reduction under biaxial strain originates from the excess energy between excitation and emission. However, it is not sufficient to explain the total depolarization. By subtracting the two slopes we eliminate contributions from the excitation energy detuning and we estimate the remaining effect emerging inherently from biaxial strain to be approximately (–0.08±0.02) %/meV (green line in Figure 5). This value suggests that applying 1% of biaxial tensile strain results in a ≈40% drop of the unstrained VP degree. We further comment on additional microscopic contributions from biaxial strain beyond energy detuning. One possibility includes intervalley hole scattering from K to Γ point as biaxial strain shifts the Γ valence band energetically closer to the K valley. However, we exclude this scenario in 1L-$WS_2$ since the energy splitting between K and Γ valence bands is hundreds of meV even for 0.45% tensile biaxial strain, making hole scattering processes energetically unfavorable[30] (see DFT calculations in Supplementary Material, section E, Figures S6a,c).

In contrast, analytical calculations[20] that take into account excitonic effects estimate that, among other processes[31], tensile biaxial strain will increase the oscillator strength (i.e. proportional to the square of the transition matrix element) and the exciton radiative broadening. This will result in a larger radiative decay rate, $\Gamma_0$, that impacts the longitudinal-transverse (L-T) exciton splitting, with $\Omega_{LT} \propto \Gamma_0 \cdot {}^{K}/_{q}$. Here, $\Omega_{LT}$ is the effective pseudospin precession frequency, $K$ is the exciton momentum and $q$ is the light wavevector at the exciton resonance frequency[32,28]. A larger $\Omega_{LT}$ will result in a shorter spin relaxation time, $\tau_s$, through the Dyakonov-Perel[28] spin relaxation, where $\tau_s^{-1} = \langle {\Omega_{LT}}^2 \tau_2 \rangle$ with $\tau_2$ being the scattering time. Thus, biaxial tensile strain could yield a decrease of the VP degree. We emphasize that biaxial strain substantially differs from uniaxial strain (i.e. $u_{xx} \neq u_{yy}$), as the latter results in a splitting of the exciton radiative doublet accompanied by a strain-induced optical anisotropy with softened optical selection rules[33] and also contributes to the exciton depolarization via the anisotropic contribution to the L-T splitting[34]. Changes in the optical matrix element only partially explain our experimental observations. Tensile biaxial strain is also expected to vary the scattering time, $\tau_2$, via modified electron (or exciton)-phonon scattering processes which in turn affect spin relaxation in 1L-$WS_2$. It has been recently shown that Λ valleys play a crucial role in the relaxation of excitons[35,36] especially in W-based monolayers where the energy difference between K and Λ conduction bands is very small[37]. In fact, the exciton landscape is very sensitive to strain with a strong involvement of Λ-valleys[37]. In agreement with previous reports[38,39], our DFT calculations in 1L-$WS_2$ (Supplementary Material, section E, Figures S6a,b) show that the energy difference between K and Λ conduction bands substantially increases even for small values of tensile biaxial strain. Consequently, one of the scattering channels (that was possible in the unstrained case) becomes suppressed resulting in a longer scattering time ($\tau_2$), further decreasing the spin relaxation time, $\tau_s$.

In summary, we experimentally investigate the effect of biaxial tensile strain on the exciton energy and degree of polarization in 1L-$WS_2$/graphite heterostructures at room temperature. We perform photoluminescence experiments, and we measure a strong exciton shift on the order of ~ -130 meV per % of strain. Under helicity-resolved near-resonant excitation conditions we measure a monotonic decrease in the VP degree under tensile biaxial strain. We distinguish different contributions to the effect. We find that 70% of the drop in the VP is due to the increase in the exciton thermalization time while the remaining 30% comes from a combination of strain-induced enhancement in the optical matrix element as well as in the suppression of exciton-phonon scattering processes. Our results provide important insights both in tuning the optoelectronic properties of 2D TMDs at room temperature and in understanding the microscopic mechanisms of exciton spin relaxation under strain.

## SUPPLEMENTARY MATERIAL

See the supplementary material for further details.

## ACKNOWLEDGEMENTS

G.Kour, and G.Kio., acknowledge funding by the Hellenic Foundation for Research and Innovation (H.F.R.I.) under the 'First Call for H.F.R.I. Research Projects to support Faculty members and Researchers and the procurement of high-cost research equipment grant' project No:

HFRI-FM17-3034. A.M., J.P., K.P., and G. Kio., acknowledge support by the project SPIVAST funded by the Foundation for Research and Technology Hellas. MMG acknowledges the RSF grant 23-12-00142 for support of theoretical analysis. B.J. and L.C. acknowledge financial support from the Research Foundation Flanders (FWO). G.Kour., I.P., X.M., L.C., E.S., and G.Kio, acknowledge support by the EU-funded DYNASTY project, ID: 101079179, under the Horizon Europe framework programme.

**Figure 1**

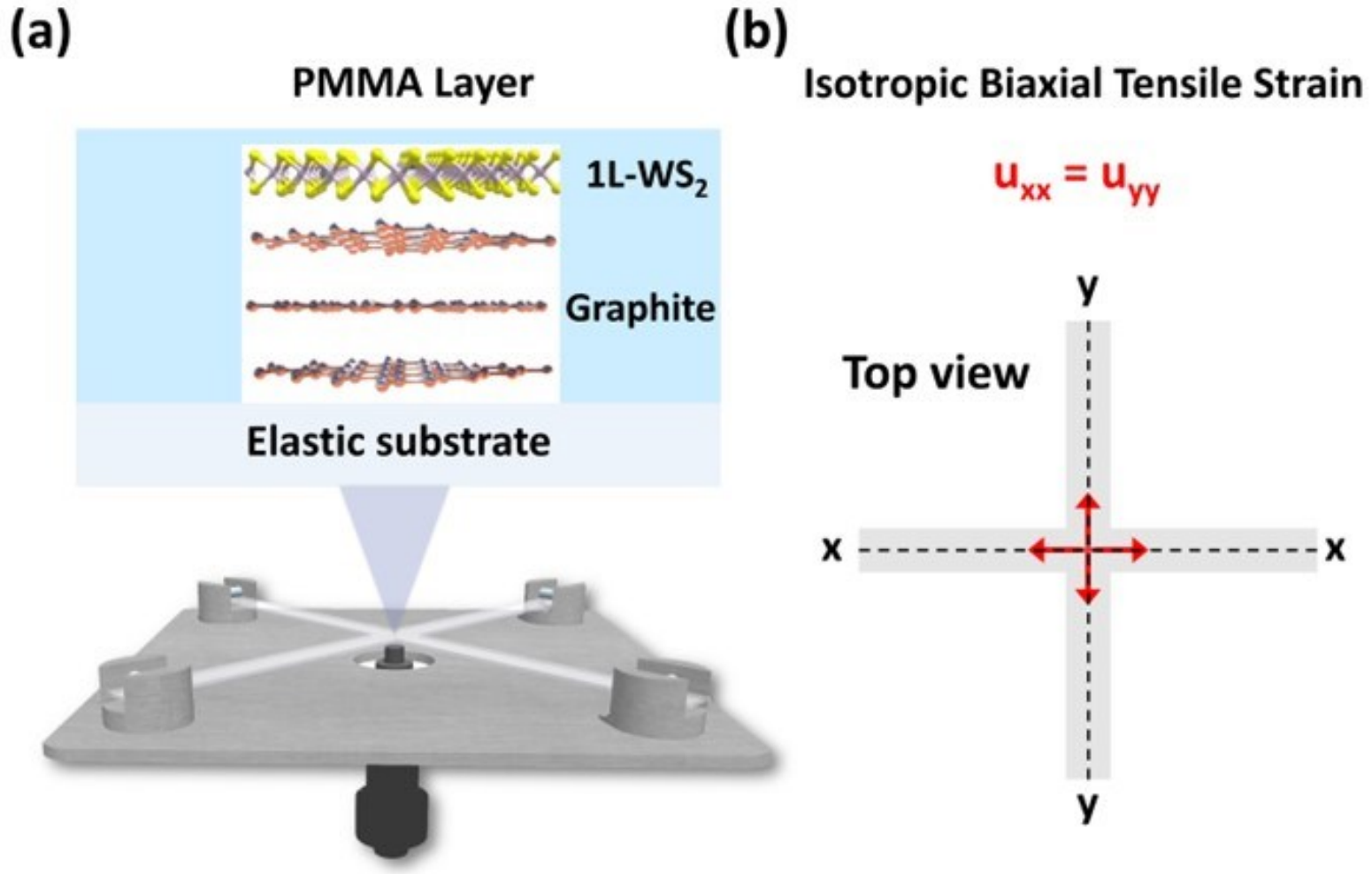

**Figure 2**

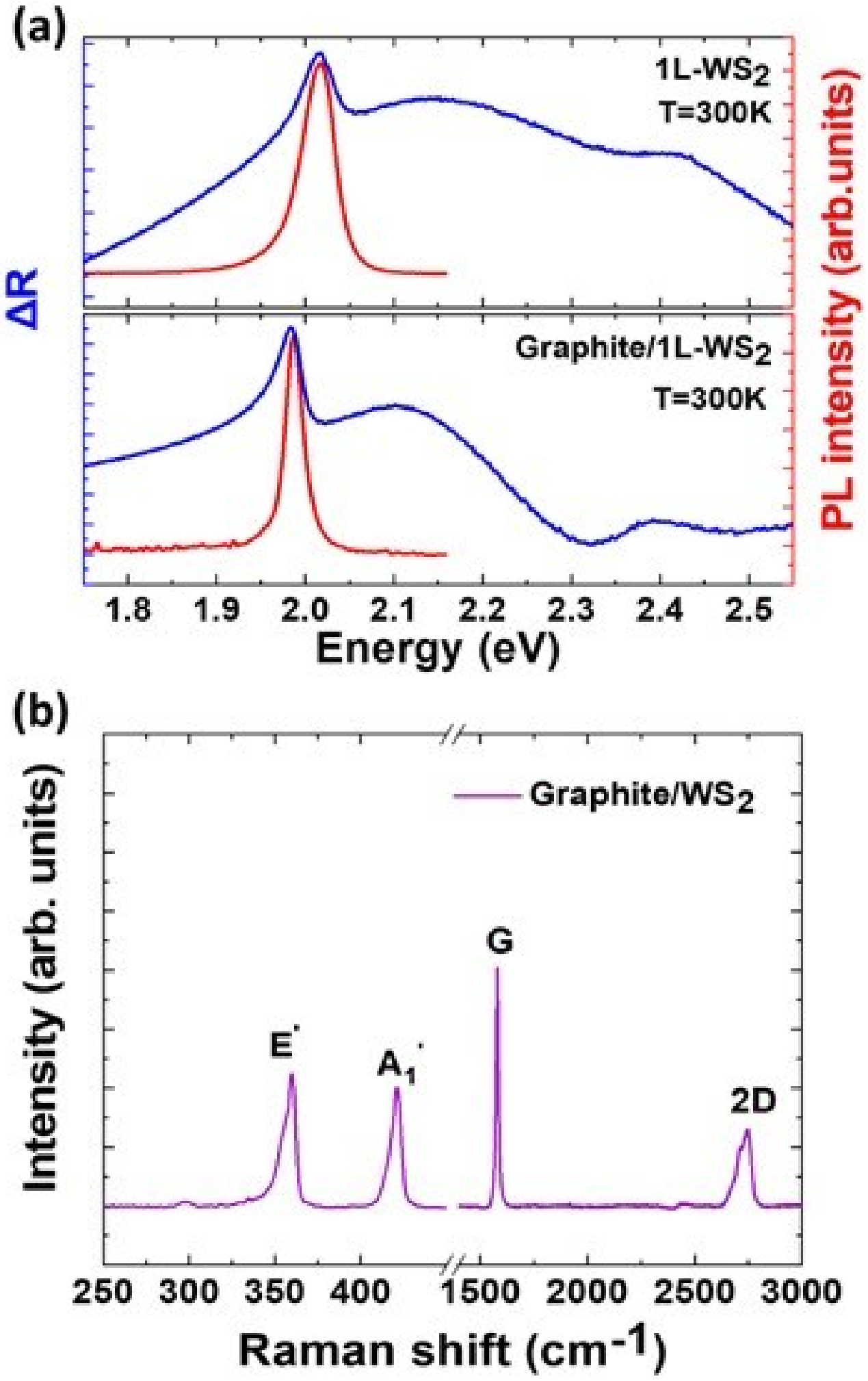

Figure 3

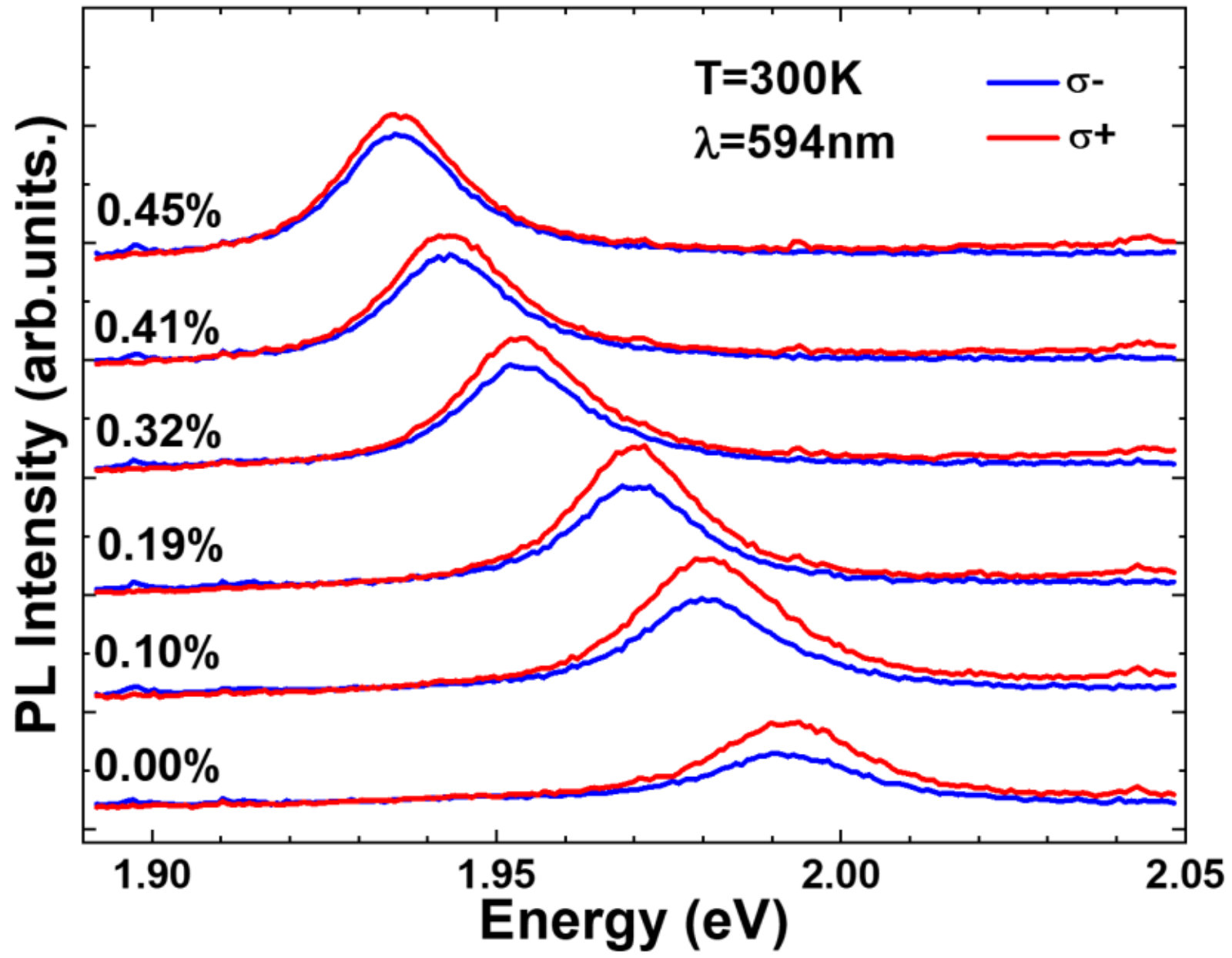

Figure 4

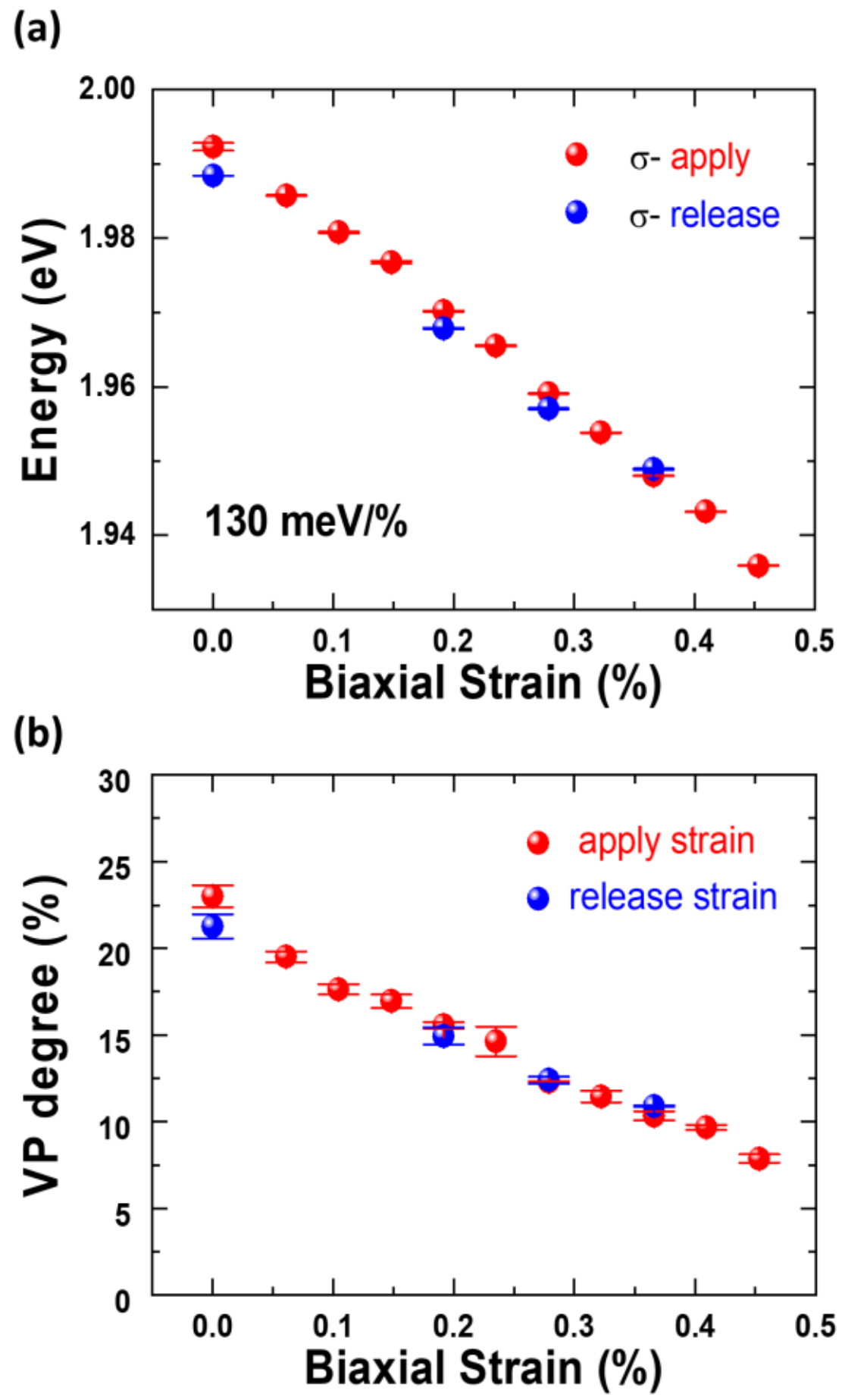

**Figure 5**

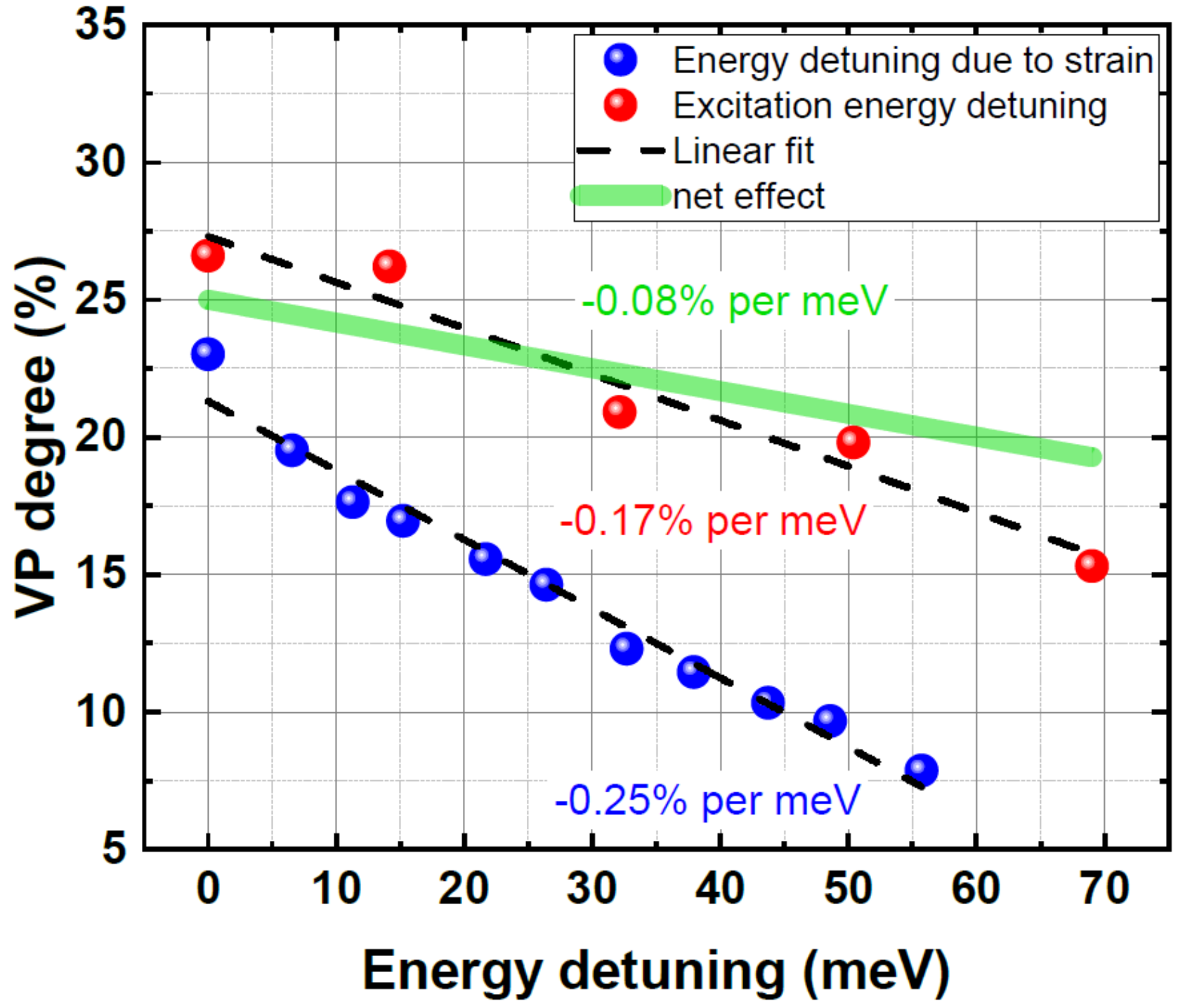

## FIGURE CAPTIONS

**Figure 1:** Schematic representation of the strain device. (a) Cross-section of the encapsulated heterostructure (Graphite/1L-$WS_2$) in PMMA layer placed on the flexible cruciform substrate for strain-related studies (b) Top view of the elastic substrate. The red arrows indicate the strain tensors applied isotropically in x and y axis.

**Figure 2:** (a) PL and differential reflectivity comparison of 1L-$WS_2$ and Graphite/1L-$WS_2$. (b) Raman spectra of the heterostructure Graphite/1L-$WS_2$.

**Figure 3:** PL spectroscopy using $\sigma^+$ polarized excitation for different values of biaxial strain. Red and blue colors correspond to $\sigma^+$ and $\sigma^-$ PL emission spectra, respectively.

**Figure 4:** (a) Energy shift of the PL emission with biaxial strain. Red and blue color correspond to data acquisition cycles of applying and releasing strain respectively. (b) Degree of valley polarization with biaxial strain. Red and blue color are linked with applying and releasing strain, respectively.

**Figure 5:** Degree of VP as a function of PL energy shift due to biaxial strain (blue spheres) and as a function of excitation energy (red spheres). Linear fits shown with black dashed lines are applied in the experimental points (spheres). The green line corresponds to the net drop of VP degree due to isotropic biaxial strain excluding the energy detuning.

## DATA AVAILABILITY

The data that support the findings of this study are available within this article and its supplementary.

# Supplementary Material

## Biaxial strain tuning of exciton energy and polarization in monolayer $WS_2$

G. Kourmoulakis[1,2†], A. Michail[3,4†], I. Paradisanos[1], X. Marie[5], M.M. Glazov[6], B. Jorissen[7], L. Covaci[7], E. Stratakis[1,8], K. Papagelis[4,9], J. Parthenios[4*], and G. Kioseoglou[1,2*]

[1]*Institute of Electronic Structure and Laser, Foundation for Research and Technology-Hellas, Heraklion, 70013, Greece*

[2]*Department of Materials Science and Technology, University of Crete, Heraklion, 70013, Greece*

[3]*Department of Physics, University of Patras, Patras, 26504, Greece*

[4]*Institute of Chemical Engineering Sciences, Foundation for Research and Technology-Hellas, Stadiou str Platani, Patras, 26504, Greece*

[5]*Universite de Toulouse, INSA-CNRS-UPS, LPCNO, 135 Av. Rangueil, 31077, Toulouse, France*

[6]*Ioffe Institute, St.-Petersburg, 194021, Russia*

[7]*University of Antwerp, Groenenborgerlaan 171, B-2020 Antwerpen, Belgium*

[8]*Department of Physics, University of Crete, Heraklion Crete 71003, Greece*

[9]*School of Physics, Dept. of Solid-State Physics, Aristotle University of Thessaloniki, Thessaloniki, 54124, Greece*

[†] These authors contributed equally

* Authors to whom any correspondence should be addressed
email: jparthen@iceht.forth.gr, gnk@materials.uoc.gr

## A. Transfer process and strain device calibration

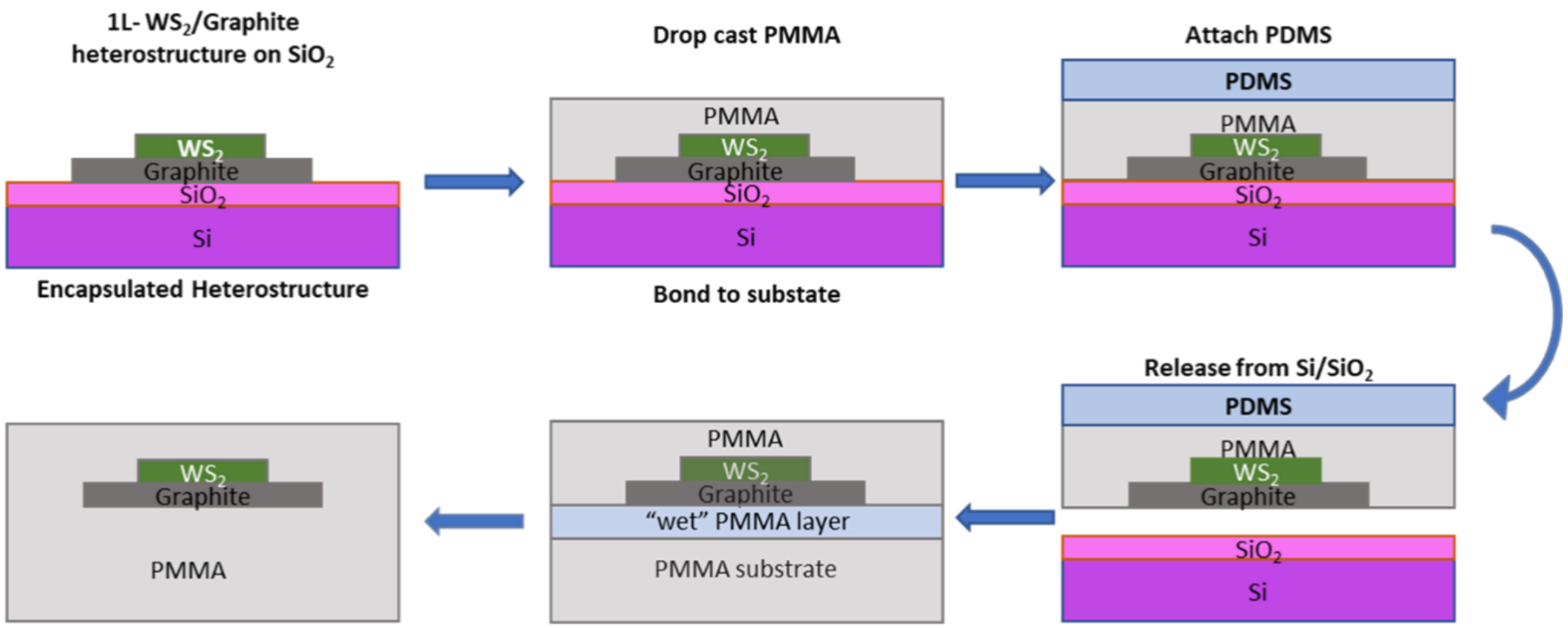

**Figure S1.** Step-by-step transfer and encapsulation of the Graphite/1L-$WS_2$ heterostructure.

A few drops of PMMA 495K A3 solution (Microchem) were casted over the heterostructure and left to dry completely. With the help of PDMS stamp, and after pinching the edges of the casted film with a pair of tweezers, the film could be removed from the Si/$SiO_2$ substrate along with the heterostructure. Then, a wet PMMA layer is spin coated at 1000 rpm for 10s on a PMMA cruciform. Before the freshly spin-coated layer dries the PDMS stamp carrying the PMMA film with the hetero structure is pressed gently on the still "wet" PMMA layers and left to bond overnight. The solvent in the fresh PMMA layer partially dissolves the thin PMMA film carrying the heterostructure and after the procedure is completed the PMMA film and the PMMA substrate bond to a single solid continuous piece, fully encapsulating the heterostructure. The encapsulation is crucial for the successful transfer of mechanical stresses to the $WS_2$ monolayer as well as the supporting graphite.

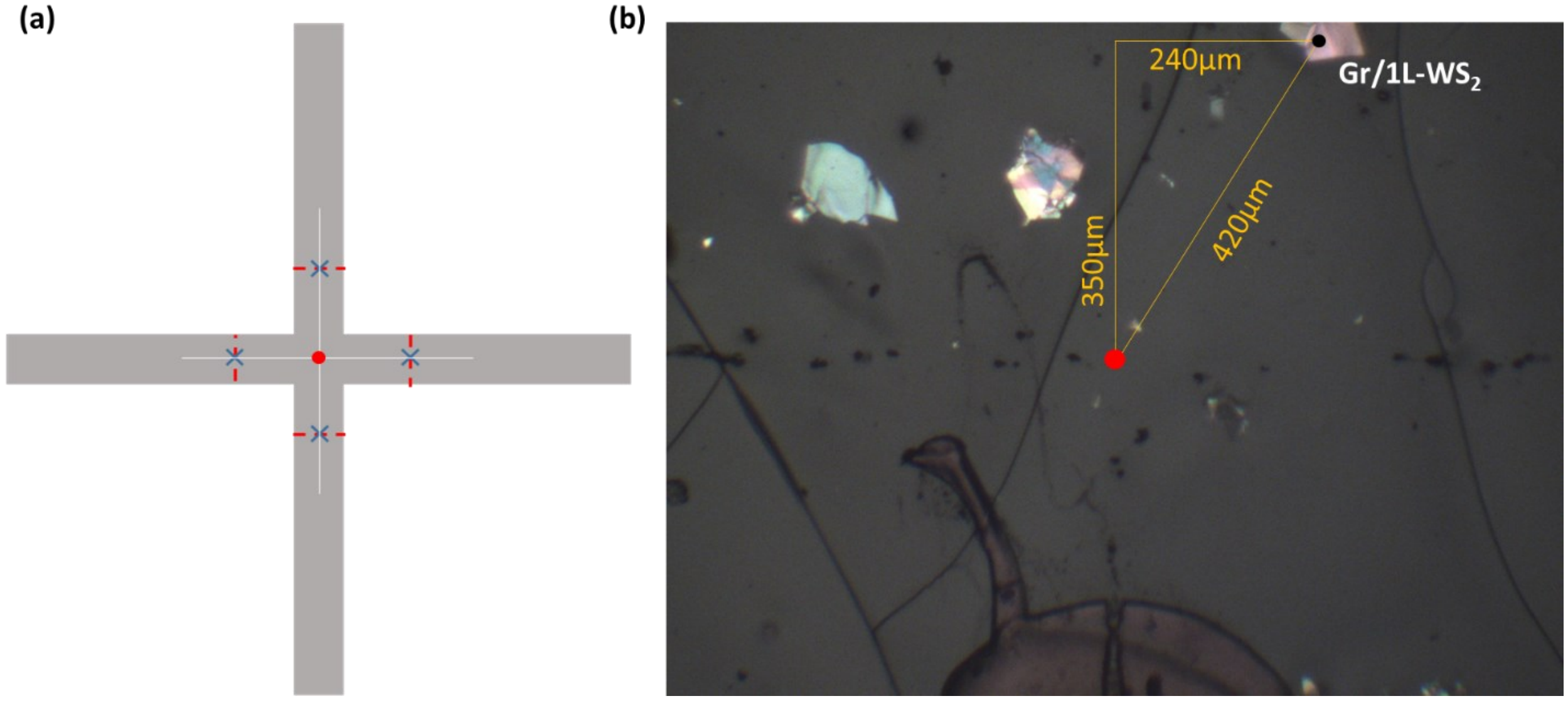

**Figure S2.** (a) Geometric image of the cruciform. The blue x marks indicate the centers of each axis, and the red dot indicates the center. (b) Optical image of the distance between the target heterostructure (marked by the black circle) and the center of the cruciform (red circle).

The cruciform was mounted on a motorized xyz stage. We measured the width of each side (red dashed lines) and marked the centers (blue x) as presented in Figure S2a. Using a femtosecond laser source, we found the center of the cruciform (red circle) by irradiating the substrate following vertical lines (white color) from the blue marks. In Figure S2b, we show the calculated distance between the heterostructure and the center of the cruciform. For the specific location of the Gr/1L-$WS_2$ heterostructure the normalized biaxial strain and the shear-to-biaxial strain ratios are 0.995 and 0.0046 respectively. The shear strain can be calculated as

$$e_s = e_{xx} - e_{yy}$$

$$e_{shear} = 0.0046 \times e_{biaxial}$$

Where, $e_{biaxial} = e_{xx} + e_{yy}$.

By performing finite element calculations (see supporting info in [1]) investigating the distribution of strain over the cruciform center, we found that the difference between the principal strains at the sample position is

$$e_{xx} - e_{yy} = 0.0046 \times (e_{xx} + e_{yy})$$

As a result, we can conclude that for the maximum value of strain applied in our studies (0.45%) the shear strain contribution is 0.002%. This magnitude of shear strain is certainly not detectable by means of Raman spectroscopy.

## *B. Optical image and spectroscopy of the heterostructure*

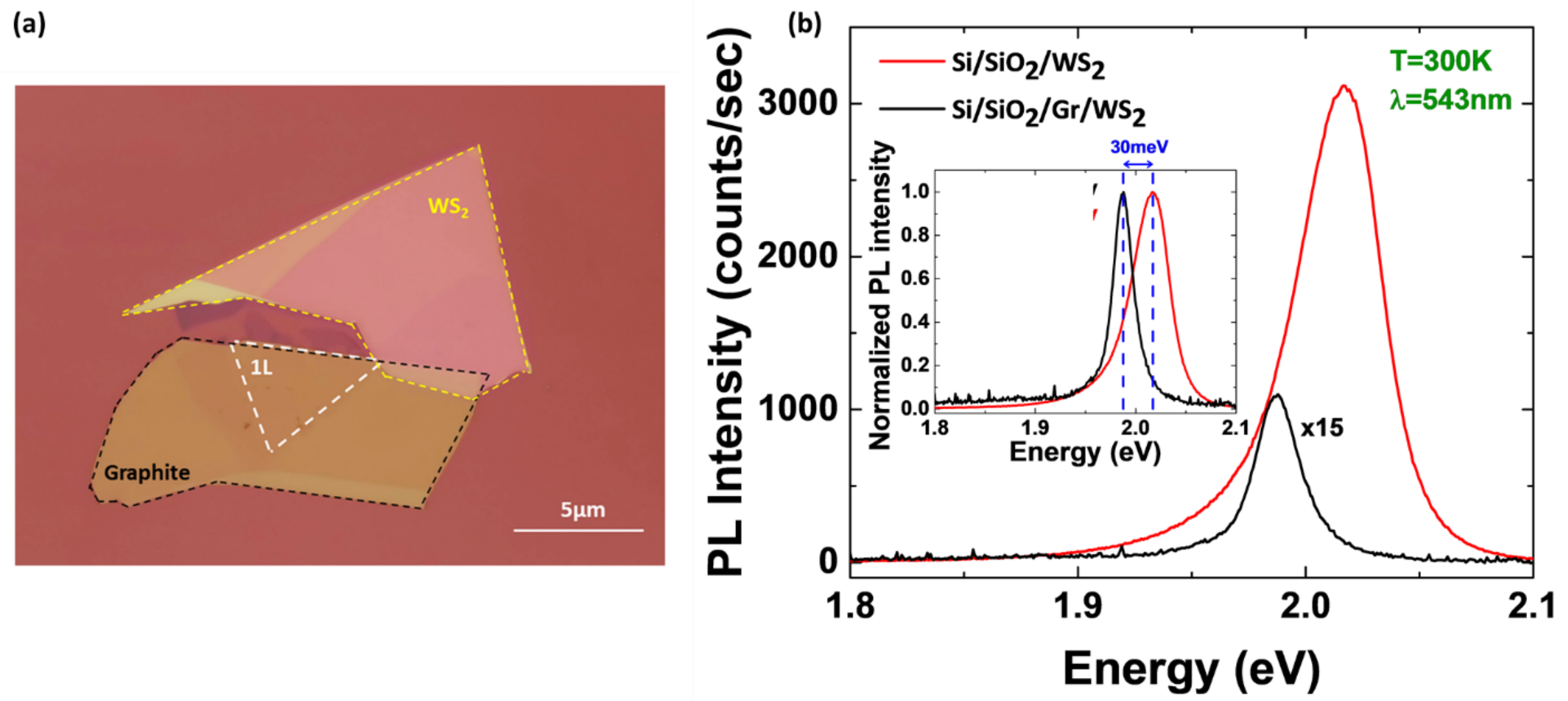

**Figure S3.** (a) Optical image of the 1L-$WS_2$/graphite heterostructure. The black, white, and yellow dashed lines denote bulk graphite, single layer $WS_2$ and bulk $WS_2$ areas respectively. (b) Room temperature PL comparison of 1L-$WS_2$ on top of Si/$SiO_2$ (red line) and bulk Graphite (black line). The inset shows the normalized PL intensity.

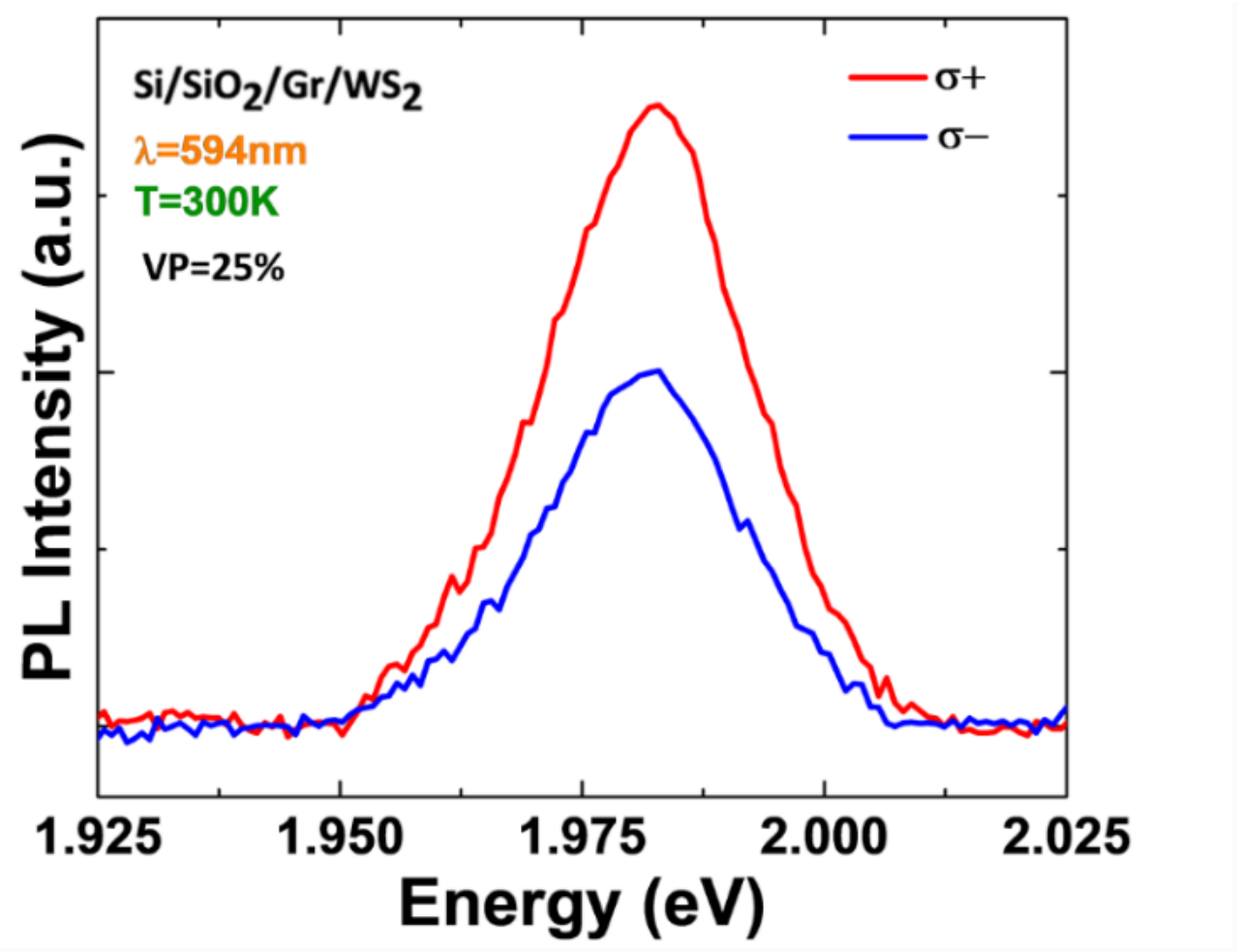

**Figure S4.** Helicity-resolved PL intensity of the Graphite/1L-$WS_2$ heterostructure with near resonance excitation (594nm).

## C. Verification of strain transfer to $WS_2$ monolayer

After successful encapsulation of the heterostructure, detailed strain dependent Raman and PL spectroscopic investigations were conducted to assess the strain-transfer efficiency to the $WS_2$ monolayer. Biaxial mechanical strain was gradually applied in steps and at each strain level a series of PL and Raman spectra were collected. In **Figure S3** (a) and (b), representative PL and Raman spectra obtained from a 1L- $WS_2$/GR heterostructure are shown. It was observed that the A exciton emission peak at 1.99 eV as well as the $2LA$, $E'$ and $A_1'$ Raman modes located at 352 $cm^{-1}$ , 357 $cm^{-1}$ and 419 $cm^{-1}$, respectively, redshift with increasing strain, as shown in Figure S3 (c) and (d). Table S-1 summarizes the results for four different points in the heterostructure. The average shift rate for the A exciton is -140(7) meV/% which is in very good agreement with the established value in the literature of about -130 meV/%[1,2] Similarly, the average shift rates of the $2LA$, $E'$ and $A_1'$ Raman modes were determined at -6.0(7), -4.5(7) and -1.7(3) $cm^{-}$/%, respectively, comparing excellently with the literature values of 6.3, 5.7 and 1.8 $cm^{-1}$/%.

**Table S-1** PL and Raman peak positions at zero strain and corresponding strain induced shift rates for four different locations of a 1L-$WS_2$/graphite heterostructure.

| | **$A$ exciton** | | **$2LA$** | | **$E'$** | | **$A_1'$** | |
|---|---|---|---|---|---|---|---|---|
| | $E_o$ | $\frac{dE}{d\varepsilon}$ | $\omega_o$ | $\frac{d\omega(2LA)}{d\varepsilon}$ | $\omega_o$ | $\frac{d\omega(E')}{d\varepsilon}$ | $\omega_o$ | $\frac{d\omega(A_1')}{d\varepsilon}$ |
| Point | (eV) | (meV/%) | ($cm^{-1}$) | ($cm^{-1}$/%) | ($cm^{-1}$) | ($cm^{-1}$/%) | ($cm^{-1}$) | ($cm^{-1}$/%) |
| 1 | 1.994 | -128(6) | 352.2 | -6.3(2) | 356.5 | -5.4(2) | 418.8 | -1.5(2) |
| 2 | 2.006 | -130(1) | 352.7 | -7.4(5) | 357.0 | -6.4(6) | 418.5 | -1.8(4) |
| 3 | 1.997 | -150(10) | 351.3 | -5(1) | 354.5 | -3(1) | 419.5 | -1.9(2) |
| 4 | 1.998 | -150(10) | 351.4 | -5(1) | 354.7 | -3(1) | 419.3 | -1.4(3) |

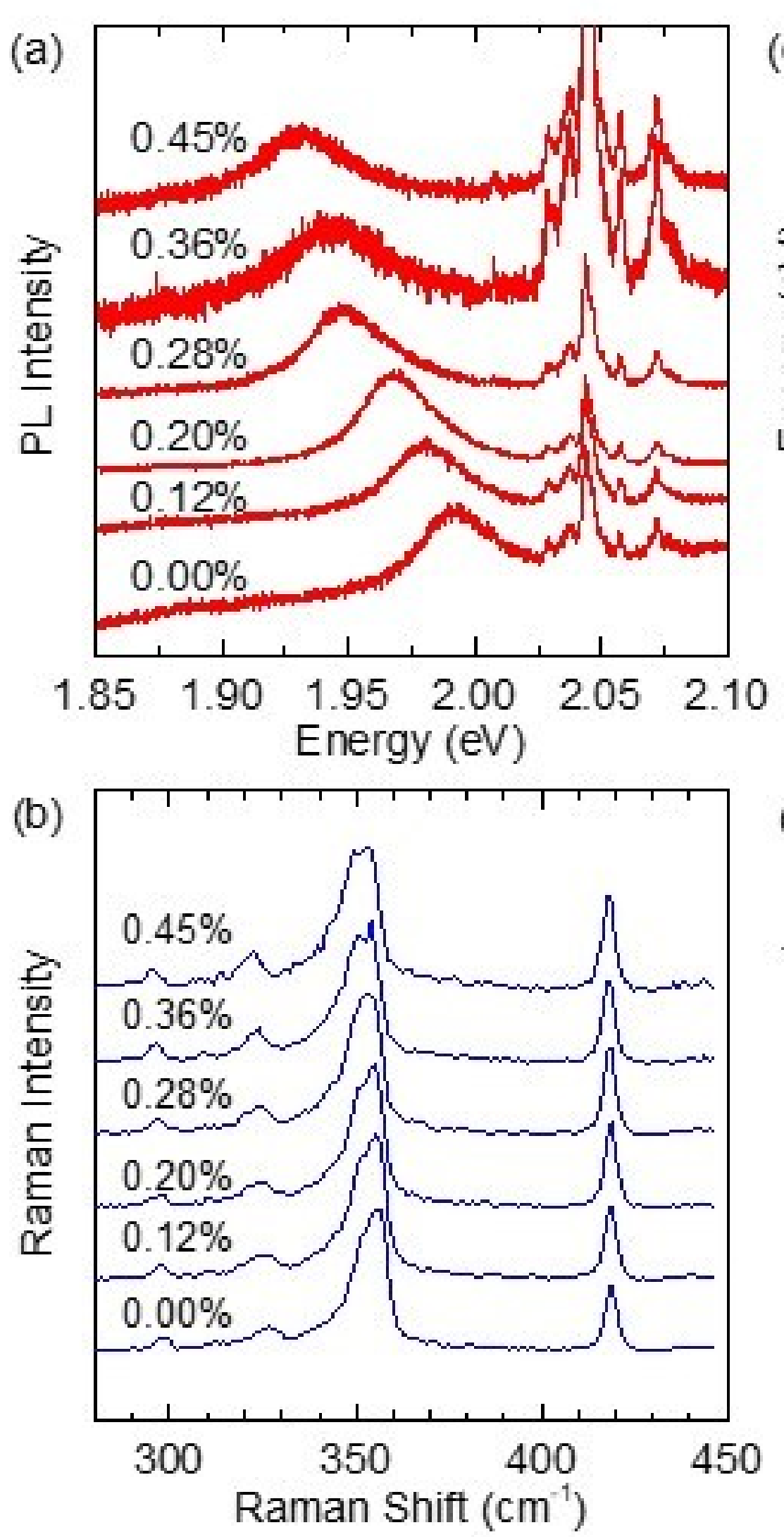

**Figure S5.** Strain dependent Raman and PL spectroscopy in a 1L - $WS_2$ / GR heterostructure. Evolution of (a) the PL and (b) Raman spectra with externally applied strain. Dependence of (c) the A exciton emission energy and (d) the $2LA$, $E'$ and $A_1'$ Raman mode frequencies on strain. ($\lambda_{exc}$ = 515 nm).

## *D. Theory of strain-induced effects on valley polarization*

Before discussing the model, we specify terminology used here. In atomically thin crystals such as graphene and TMD monolayers, the most important is the in-plane strain components $u_{xx}$ and $u_{yy}$ of the strain tensor:

$$u_{\alpha\beta} = \frac{1}{2}\left(\frac{\partial u_\alpha}{\partial x_\beta} + \frac{\partial u_\beta}{\partial x_\alpha} + \frac{\partial u_\gamma}{\partial x_\alpha}\frac{\partial u_\gamma}{\partial x_\beta}\right).$$

Here $\alpha, \beta, \gamma$ running through $x, y, z$ denote Cartesian components, $u_\alpha$ are the components of the displacement vector, and $x_\alpha$ are the components of the position vector. We call the in-plane strain with $u_{xx} = u_{yy}$ as *biaxial* strain, and the anisotropic in-plane strain with $u_{xx} \neq u_{yy}$ as the *uniaxial* strain.

There are two basic mechanisms of the strain effect on the exciton valley polarization in monolayer transition metal dichalcogenides. First one, the band-structure mechanism, is related to the strain-induced modification of the band structure and optical selection rules for the interband transitions. Second one, the kinetic mechanism, is related to the valley polarization dynamics.

We start the analysis with the *band-structure mechanism*. For instance, the anisotropic uniaxial strain in the monolayer plane induces the optical anisotropy of the monolayer and breaks the strict chiral selection rules already at the K and K' points of the Brillouin zone[3]. The biaxial strain does not break the three-fold rotation symmetry and preserves clean chiral selection rules, where $\sigma^+$photon induces the transition at the K point while the $\sigma^-$ photon activates the transition at the K' point. However, the biaxial strain affects the exciton transition energy $E_X = E_g - E_b$, with $E_g$being the band gap and $E_b$ being the exciton binding energy, via the modification of the band gap and binding energy. Both these quantities vary $\propto u_{xx} + u_{yy}$. The variation of the band gap is related to the combination of the conduction and valence band deformation potentials[4,5]. The exciton binding energy can be affected, e.g., via the strain-induced modification of the effective masses of electron and hole forming an exciton. Experimentally, the polarization of the neutral exciton emission is monitored following excitation with the excitation energy $E_{exc} > E_X$. The optical selection rules away from the K, K' points are not chiral anymore due to the band mixing effects also known as band non-parabolicity. Calculations reported in[6] show that the non-parabolicity of the band indeed reduces the valley polarization degree. This reduction is controlled by a small, in our experiments, parameter $(E_{exc} - E_X)/E_g \ll 1$. This effect alone cannot explain the energy detuning dependence of the valley polarization reported in Fig. 5 of the main text.

To explain the experiment, we need to invoke the *kinetic mechanism*. The exciton pseudospin dynamics in the presence of uniaxial (anisotropic) strain has been studied theoretically in detail in Ref.[7]. Here we focus on the biaxial strain case. It is commonly accepted that the exciton pseudospin dynamics is controlled by the Dyakonov-Perel'-like mechanism[8,9]: The exciton pseudospin $\boldsymbol{S}$ precesses in the effective magnetic field with the precession frequency $\boldsymbol{\Omega_K}$ which depends on the exciton wavevector $\boldsymbol{K}$. Scattering processes characterized by the correlation time $\tau_2$ randomly break this precession and result in the exciton decoherence and depolarization. The depolarization rate in the strong scattering regime, $\Omega_K \tau_2 \ll 1$, is given by[7–9]

$$\frac{1}{\tau_s(\epsilon)} = \langle \Omega_K^2 \tau_2 \rangle,$$

where the angular brackets denote the averaging over the directions of $K$ and $\epsilon = \hbar^2 K^2/(2M)$ is the exciton kinetic energy with $M$ being its effective mass. Since, for excitons in monolayer semiconductors[6,10,11]

$$\Omega_K \approx \frac{\Gamma_0 K}{\kappa_{\text{eff}} q},$$

where $\Gamma_0$ is the exciton radiative decay rate, $\kappa_{\text{eff}}$ is the effective screening constant of the long-range exchange interaction[11] and $q$ is the light wavevector at the frequency of exciton resonance, the pseudospin relaxation time $\tau_s$ strongly depends on the exciton kinetic energy. For example, for quasi-elastic acoustic phonon scattering $\tau_2$is energy independent (it depends on the temperature via the phonon occupancies, see Ref.[5] for details), $\tau_s(\epsilon) \propto \epsilon^{-1}$. Hence, an increase of the excitation energy results in higher exciton kinetic energy and faster depolarization. As a result, the observed valley polarization degree reduces.

Following Refs.[12,13] we present the depolarization factor related to the spin relaxation in the course energy relaxation as

$$\zeta = \exp\left( - \int_0^\Delta \frac{\tau_s(\epsilon)}{\tau_\epsilon} \frac{d\epsilon}{\epsilon} \right).$$

Here $\Delta = E_{exc} - E_X$ is the energy detuning and $\tau_\epsilon$ is the energy relaxation time. In the kinetic mechanism, the biaxial strain affects all three key parameters in this expression:

- Energy detuning $\Delta$ via the exciton energy $E_X$ as discussed above;
- Spin relaxation time $\tau_s(\epsilon)$ both via variation of $\Omega_K$ and $\tau_2$; and
- Energy relaxation time $\tau_\epsilon$.

All these effects results in

$$\zeta \propto u_{xx} + u_{yy}.$$

Our experiments with the tensile strain demonstrate the reduction of polarization. Hence, we expect that the compressive biaxial strain results in the enhancement of polarization. In any case, the effects described here are linear in the strain. It is instructive to compare this result with the expectation for the uniaxial (anisotropic) strain, where the key effect is the exciton depolarization via anisotropic $K$-independent contribution to $\Omega_K$, see Refs.[3,7] for details. In that case the depolarization is described by a Hanle-like profile and for not too high strains

$$\zeta \propto \left(u_{xx} - u_{yy}\right)^2,$$

i.e., the uniaxial strain effect is quadratic in the strain.

## *E. DFT calculations of band structure as a function of strain*

The DFT-PBE[13] calculations in this work were performed using Abinit[14] with the norm-conserving full-relativistic pseudopotentials from PseudoDojo[15,16]. The DFT calculations were converged with an energy cutoff 40Ha on a 10x10x1 reciprocal space grid and a convergence criterium of $10^{-4}$Ha. The monolayers were separated with a 2.5 nm vacuum region. The energy levels are aligned according to the maximum of the valence bands at the Γ-point.

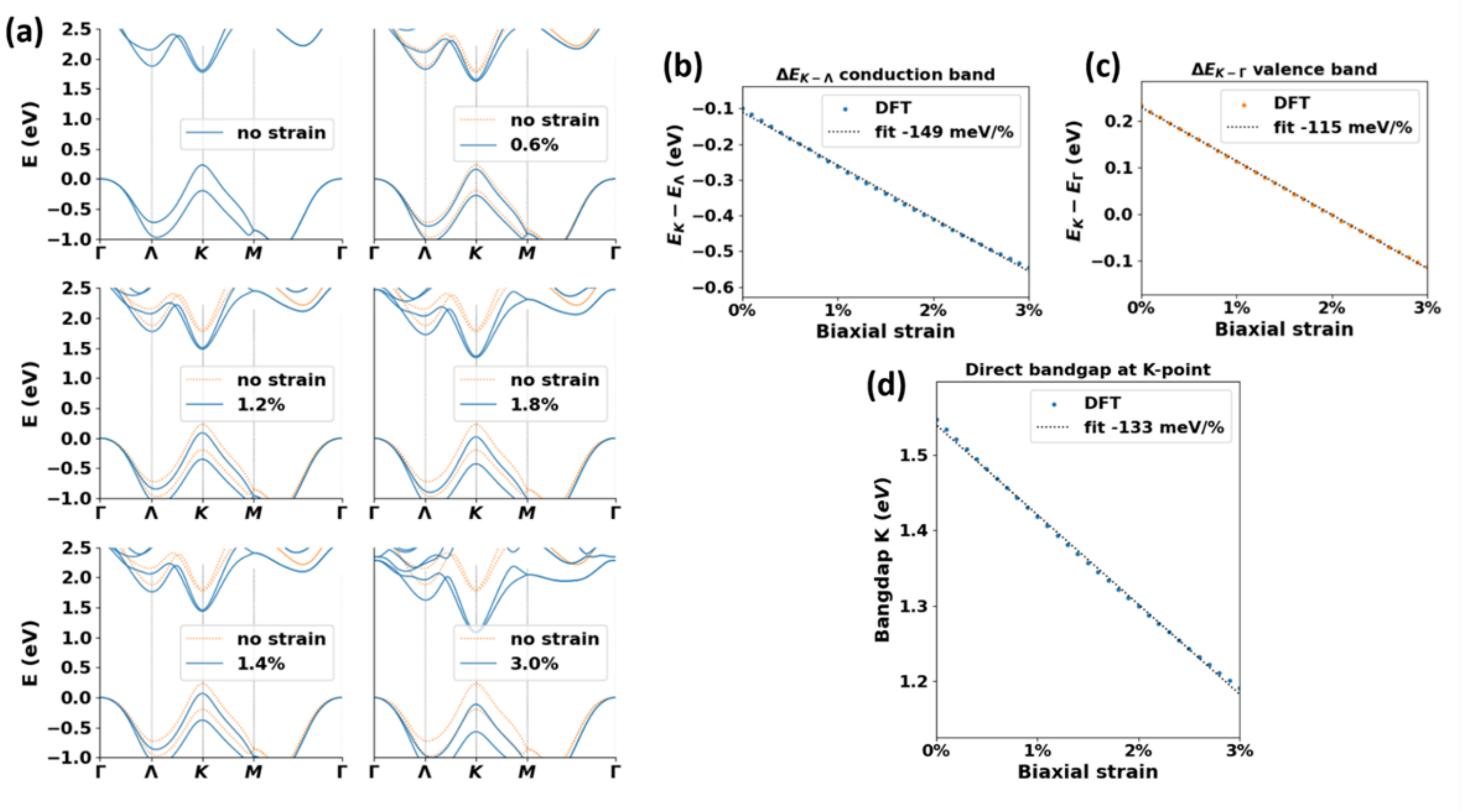

**Figure S6.** (a) Band structure DFT calculations of 1L-$WS_2$ as a function of biaxial strain. (b) Energy difference between K and Λ points of the conduction band as a function of strain. (c) Energy difference of the K and Γ points of the valence band as a function of strain. (d) Direct bandgap at the K point as a function of strain.